\documentclass[12pt,preprint]{aastex}

\newcommand{\spat}{Blackett Laboratory, Imperial College London, 
South Kensington, London SW7 2BW, UK}
\newcommand{\mssl}{Mullard Space Science Laboratory, University College 
London, Holmbury St. Mary, Dorking, Surrey RH5 6NT, UK}
\newcommand{\icmaths}{Department of Mathematics, Imperial College London, 
South Kensington, London SW7 2BW, UK}

\slugcomment{Submitted to Astrophys. J. Lett, 14 April 2005;\\ revised 23 May 2005; accepted 27 May 2005}

\shorttitle{Observations of SGR 1086-20 $\gamma$-ray flare}
\shortauthors{Schwartz et al.}

\begin{document}
\title{The $\gamma$-ray giant flare from SGR1806-20: Evidence for crustal 
cracking via initial timescales}

\author{Steven J. Schwartz\altaffilmark{1}}\email{s.schwartz@imperial.ac.uk}
\and\author{Silvia Zane\altaffilmark{2}}\email{sz@mssl.ucl.ac.uk}
\and\author{Robert J. Wilson\altaffilmark{2}}\email{rjw@mssl.ucl.ac.uk}
\and\author{Frank P. Pijpers\altaffilmark{1}}\email{f.pijpers@imperial.ac.uk}
\and\author{Daniel R. Moore\altaffilmark{3}}\email{dan.moore@imperial.ac.uk}
\and\author{D. O. Kataria\altaffilmark{2}}\email{dok@mssl.ucl.ac.uk}
\and\author{Timothy S. Horbury\altaffilmark{1}}\email{t.horbury@imperial.ac.uk}
\and\author{Andrew N. Fazakerley\altaffilmark{2}}\email{anf@mssl.ucl.ac.uk}
\and\author{Peter J. Cargill\altaffilmark{1}}\email{p.cargill@imperial.ac.uk}

\altaffiltext{1}{\spat}
\altaffiltext{2}{\mssl}
\altaffiltext{3}{\icmaths}

\begin{abstract} Soft $\gamma$-ray repeaters (SGRs) are neutron stars which
emit short ($\lesssim 1$s) and energetic ($\lesssim 10^{42}$erg s$^{-1}$)
bursts of soft $\gamma$-rays. Only 4 of them are currently known. Occasionally,
SGRs have been observed to emit much more energetic ``giant flares'' ($\sim
10^{44}-10^{45}$erg s$^{-1}$).  These are  exceptional and rare events. We
report here on serendipitous observations of the intense $\gamma$-ray flare
from SGR 1806-20 that occured on 27 December 2004. Unique data from the Cluster
and Double Star-2 satellites, designed to study the Earth's magnetosphere,
provide the first observational evidence of three separate timescales  within
the early (first 100ms) phases of this class of events. These observations
reveal that, in addition to the initial very steep ($<$0.25ms) X-ray
onset, there is firstly a 4.9ms exponential rise timescale 
followed by a continued  exponential rise in intensity on a timescale of 70ms.
These three timescales are a prominent feature of current theoretical
models, including the timescale (several ms) for
fracture propagation in the crust of the neutron star.

\end{abstract}


\section{Introduction}

During the quiescent state (i.e., outside bursts events) Soft $\gamma$-ray
Repeaters (SGRs) are detected as
persistent X-ray emitters with a luminosity of $\sim 10^{35}$erg s$^{-1}$. 
Several characteristics of SGRs, including their bursting  activity, are often
explained in the context of the ``magnetar''
model \citep{duncan+thompson92,thompson+duncan95,thompson++02}.  Magnetars are 
neutron stars whose X-ray emission is attributed to the presence of an
ultra-strong magnetic field ($\sim10^{14}-10^{15}$Gauss). The frequent short
bursts are associated with small cracks in the neutron star crust, driven 
by
magnetic diffusion \citep{thompson+duncan95}. Alternatively, because 
of their short (submillisecond) time scale,  they may follow  a 
sudden loss of magnetic equilibrium through the development of a tearing
instability \citep{lyutikov02,lyutikov03}. Giant flares, acting on a much 
longer time scale, should involve not only the above phenomena but also 
global rearrangements of the magnetic field in the interior and 
magnetosphere of the star.

Observations made on  2004 December 27 showed that {SGR} 1806-20 had experienced
an exceptionally powerful giant flare, lasting $\sim 0.25$s 
\citep{borkowsky++04,hurley++04,mazets++04,palmer++05,mereghetti++05}.  For the
first 200ms it saturated almost all instruments on satellites equipped to
observe $\gamma$-rays.  Instruments on board the {GEOTAIL} spacecraft provided
unique measurements \citep{terasawa++05} of the hard X-ray intensity throughout
the event.   The energy associated with $\gamma$-rays  above 50 keV was $\sim 5
\times 10^{46}$ ergs (assuming a distance of 15 kpc and that it radiates
isotropically), hundreds of times brighter than the giant flares previously
observed from other SGRs.  Following the event, a radio afterglow has been
detected \citep{cameron+kulkarni05,gaensler++05}. Several follow-on studies
\citep[e.g.][]{hurley++05,israel++05} have concentrated on the characteristics
of the extended tail of the flare.

In this letter we report observations of the initial flare rise and decay of
SGR 1806-20. This phase was recorded by the thermal electron 
detectors \citep{johnstone++97} onboard two of the four Cluster spaceraft, and
also by an identical instrument on the  Chinese Double Star polar spacecraft,
TC-2 \citep{fazakerley++05}. The data reported here are unsaturated, have a
slightly higher temporal resolution (4ms as opposed to 5.48ms) than that of
similar measurements taken by the GEOTAIL spacecraft \citep{terasawa++05}, and,
critically, captured the initial flare activity without interruption during the
very  steep rise through and beyond the maximum in intensity. This dataset
sheds unique information on the most intense period of flare activity.

\section{Method}
\label{methods}

We have drawn on data from the PEACE instruments on Cluster-2
(C2, not shown below), Cluster-4 (C4), and Double Star Polar Orbiter (TC-2). Altogether, these
spacecraft carry 5 nearly identical instruments, although the naturally
occuring electrons, which the instruments are designed to detect, confuse the
flare response on two of them (the Low Energy Electron Analysers on C2 and C4).
As the remaining sensors provide the same coverage, with less interference, we
concentrate on those. The Cluster and TC-2 spacecraft were at approximately  
42,000km and 26,000km altitudes respectively and had unobstructed sight lines
to the source, which was  $5.3^\circ$ from the Earth-Sun
line \citep{mazets++05}. Additionally, all the sensors used in the present work
were not obstructed by other parts of the spacecraft during the initial flare
spike.

The PEACE detector onboard TC-2 sweeps through a full range of electron energies
every 121ms followed by an 8ms gap. Each sweep is divided into 30 energy channels, measured
sequentially and returned in the telemetry stream, leading to a 4ms cadence in
count rate values. At the time of the flare, the instrument was operating in a mode in which
the lowest 14 channels were not
returned, leading to a sequence of 16 measurements at 4ms resolution (64ms in total) followed by a gap of
56+8ms. The response of the instrument to X- and $\gamma$-radiation has not been
calibrated, but the count rates are due to direct stimulation of the
micro-channel-plate detectors at the receiving end of the electron
optics. As such, it is independent of the particular energy
channel and  proportional to the incident flux, but dependent on the
photon energy.

By contrast, the PEACE detectors onboard the other spacecraft (C2 and C4) were operating in a different mode. Firstly, neighbouring energy channels were summed pairwise, leading to an 8ms cadence within a single sweep. Moreover, neighbouring sweeps were summed pairwise. Thus an individual measurement is the sum of an 8ms sample with another 8ms sample taken 125ms later.
While there is no
possibility to deconvolve this data, we are able to use the steep rise and
decay of the flare to our advantage by assuming that each datum is dominated by
the 8ms sample located in time closest to the peak in flare intensity. The data show changes in
count rate of an order of magnitude per sweep. This enables us to assign those
values, with a 10\% error in counts, to a unique time. However, near the peak
itself, where the counting rates vary less rapidly, this technique leads to a
larger error. Additionally, on Cluster the count rate due to natural eletrons
increases toward the end of each 125ms sweep. In this manner we reconstruct, with some unavoidable error, a time series of 8ms samples in 128ms blocks with 125ms gaps between blocks.

The response of the instruments on Cluster falls off more rapidly
with time beyond the flare peak than on TC-2. This is most likely due
to the changing instrumental response as the spacecraft spins, as described
below; it may also be indicative of changes in the
spectral characteristics of the source \citep{mazets++05}. We have
drawn most of our conclusions from the higher resolution data on TC-2, which
follows more of the event and with less natural electron interference, and used
the Cluster data to fill in information about the shape of the light curve
during a single 56ms period when TC-2 did not return data. 

We have not attempted a determination of the absolute $\gamma$-ray flux during
the event. Apart from the lack of normal ground calibrations (the instruments
were not designed to measure $\gamma$-rays), the response depends on the
orientation of the instrument. This is due to differences in shielding added to
reduce background radiation effects on the measurements of in situ electrons,
to the partial obscurations caused by other instruments, fuel tanks, and
spacecraft body, to differences in the intrinsic detector response, and to the
operational parameters of the detector. Some of these will vary with spacecraft
spin, which has a period $\sim 4$s on both C4 and TC-2. Both TC-2 and the
sensor used here from C4 were relatively unobscured during the peak in
intensity, but had turned through $\sim 35^\circ$ by the time of a secondary
peak some 390ms later, shown below. Thus while local features are
well-represented, the relative intensities of more widely separated features
should be used with caution.

\section{Results}

Figure~\ref{dslightcurve} shows the count-rate of the TC-2 electron detector
over a 1.45 second interval. Zero time corresponds to the peak in count rate.
We also show the previously reported GEOTAIL data \citep{terasawa++05}, aligned
in time.  Where  data is available from both spacecraft the detailed features
match extremely well; in particular the steep rise  and peak are evident.
Following the first 64ms gap in the TC-2 data (see Section~\ref{methods}), a shoulder
appeared followed by a steeper decline. Two groups of data points later, at $t
\sim 390$ms, a secondary, noisy peak can be seen (enlarged in the inset), which
coincides in time and shape  with that seen by the BAT instrument on SWIFT
\citep[Supplementary Figure\ 2 of][]{palmer++05} and  also by GEOTAIL
\citep{terasawa++05}. Count rates return to their ambient levels ($\lesssim
10$) at a time 600ms beyond the peak in the count rate.  There is no doubt that
the large, but unsaturated, count rates shown in Figure~\ref{dslightcurve} are
signatures of the $\gamma$-ray giant flare emitted by SGR 1806-20. 

There is some hint in the Cluster data (not shown) of a  signal 3.5--4
seconds after the main spike. This coincides with a broad feature seen
in the BAT data  which marked the
return of the periodic  tail pulses at the same period ($\sim$7.56s) as in the
pre-flare observations.

\begin{figure}
\plotone{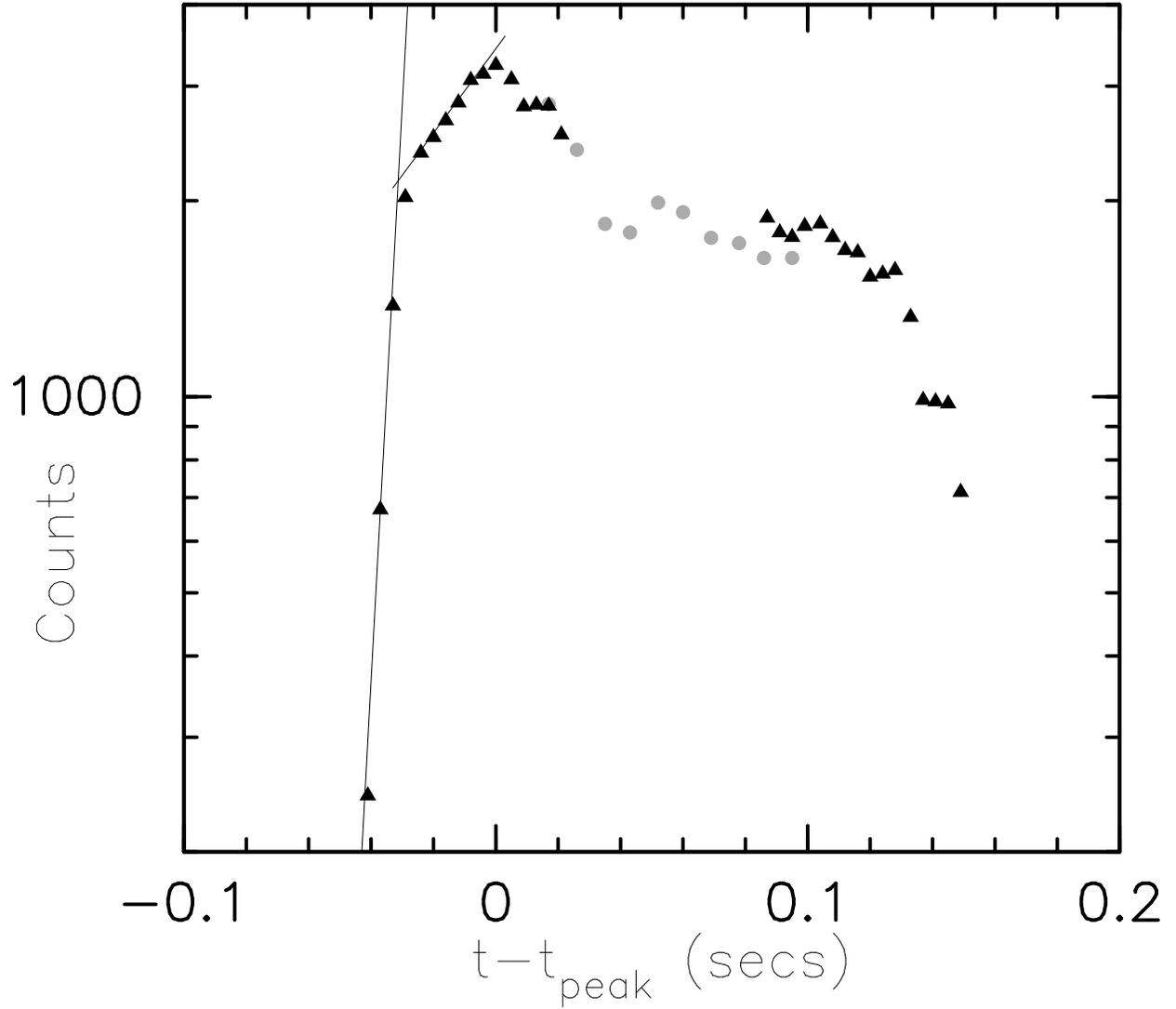}
\caption{Combined Double Star (TC-2) (triangles) and Cluster-4 HEEA
(gray circles) count rates. The data have been shifted in time to align features near
the peak
count rates. The  C4 data have been reduced by a factor 2.9; only uncomtaminated
and weakly convoluted C4 points are shown (see Section~\ref{methods}). 
Solid lines show exponential fits to the steepest TC-2 rise,
and also to the TC-2 determination of the period leading to the
main peak.
\label{composite}}
\end{figure}

In order to fill in the key time interval near the spike, we show in
Figure~\ref{composite} the combined data from both C4  and TC-2. 
The combined set reveals  a dip after $t=0$s followed by an increase  leading
to the shoulder seen in the TC-2 data. Such ``repeated energy-injections'' on a
100ms timescale were reported in the GEOTAIL data \citep{terasawa++05} as can
be seen in Figure~\ref{dslightcurve}. This 100ms timescale is a feature of all three known giant flares \citep{cline++80,feroci++01}. The C4 data prior to the peak show
evidence of time-convolutions described in Section~\ref{methods} and have been
omitted from Figure~\ref{composite}. Accordingly, we focus mainly on the TC-2
data in what follows.

The steep initial rise is well fit by an exponential function (shown as the
solid line) and has an $e$-folding time of 4.9ms. This is an order of magnitude
longer than the 0.3ms timescales \citet{palmer++05} find in the
detailed leading edge (prior to saturation of the BAT instrument). Later, 24ms
before the peak intensity was reached, the increase slowed to an exponential
rise with an $e$-folding time of 67ms, also shown as a solid line in the
figure. The previously unresolved $\sim$5ms $e$-folding rise marks the transition to a
timescale comparable to the 100ms timescale of the overall main peak and
subsequent energy injection(s). 

\section{Discussion and Conclusions}

\citet{thompson++02} recently proposed a scenario in
which the magnetars (AXPs and SGRs) differ from standard radio pulsars
since their magnetic field is globally twisted inside the star, up to a
strength of about 10 times the external dipole. At intervals it can
twist up the external field and propagate outward through Alfv\`en waves. 
The resulting stresses built up in the
neutron star crust and the magnetic footpoint movements can lead to
crustal fractures, glitching activity and flare emission \citep{ruderman91,ruderman++98}.
 In this scenario the giant flare
is produced when the crust is unable to respond (quasi)plastically any
more to the imparted stress and finally cracks 
\citep{thompson+duncan01,thompson++02}; the flare emission 
and the crustal fracturing can be in turn accompained by a simplification 
of the exterior magnetic field with a (partial) untwisting of the global 
magnetosphere.

We note here that all three of the timescales observed during the giant 
flare emitted by SGR~1806-20 can be explained in this scenario. The
initial sub-ms timescale revealed by BAT is typical of reconnection
processes in the external magnetosphere, where the low density gives rise
to short Alfv\`en times. The longer $\sim 100$ms timescale is related to
i) the overall duration of giant flares, ii) the final part of the rise
observed here and iii)  the repeated intensity rises at $\sim 100$ms
intervals observed with GEOTAIL in the first 200ms after the peak. This
time scale is comparable to the time for a $10^{15}$G magnetic field to
rearrange material in the deep crust and core of the neutron star, at an
internal Alfv\`en time, and appears as a rather regular feature in the
lightcurves of giant flares \citep{cline++80}.  The intermediate $\sim 5$ms
time is naturally explained if the rising time is limited by propagation
of a triggering fracture of size $\sim 5$km given the theoretical
expectation $\ell \approx 4$km($t_{rise}/4$ms) \citep{thompson+duncan01}.

An alternative but extreme possibility is that the giant flare is a purely magnetospheric process in which a $\sim 10$ms timescale is that which is necessary for the tearing instability to initiate large-scale magnetospheric reconnection \citep{lyutikov02,lyutikov03}. However \citet{lyutikov03} concedes that the crustal fracture model is more likely to explain the extended post-flare activity. Both perspectives require an initial fast process on a magnetospheric Alfv\'en crossing time to explain the very steep, sub-ms rise. Thus, while the 5ms timescale reported here could be related to a tearing instability in the magnetosphere,
it seems more probable that it is the first of a number of crustal events. This may not be the case for short bursts seen more frequently from SGRs, for which tearing could play an important role.

%An alternative but extreme possibility is that the giant flare was 
%caused by a global untwisting or explosive reconnection in the external 
%magnetosphere. Although this is consistent with the  energetics and with a %rise
%time of a few seconds (related to   the development of a tearing
%instability \citep{lyutikov02,lyutikov03}, the  whole process must then
%proceed on much faster timescales ($\sim 0.2$~ms)  without regard to
%the temporal scales typical of the stellar interior.  Therefore, although this
%model appears as a viable option for explaining  the short bursts observed
%from SGRs, a major problem arises for  application to giant flares of  a few
%hundred ms duration. 

Recently, tens of Hz 
Quasi-Periodic Oscillations have been detected in the tail of the 
event \citep{israel++05}. These modes are likely to be 
associated to global seismic oscillations. In particular, the large crustal 
fracturing inferred by us can easily excite toroidal modes with characteristic frequencies in 
the observed range. 

The TC-2 data thus provide the clue to the missing link between
the interior magnetic processes and their external consequences and probe
directly the crustal properties.
Cluster and Double Star were designed to study the various boundary layers of
the Earth's magnetosphere, including the physics of magnetic reconnection. Such
boundary layer physics has application throughout the astrophysical plasma
universe, and it is therefore appropriate that these missions contribute in a more direct way to the study of
magnetic reorganisation in an astrophysical object outside the solar system.

\begin{figure} 
\plotone{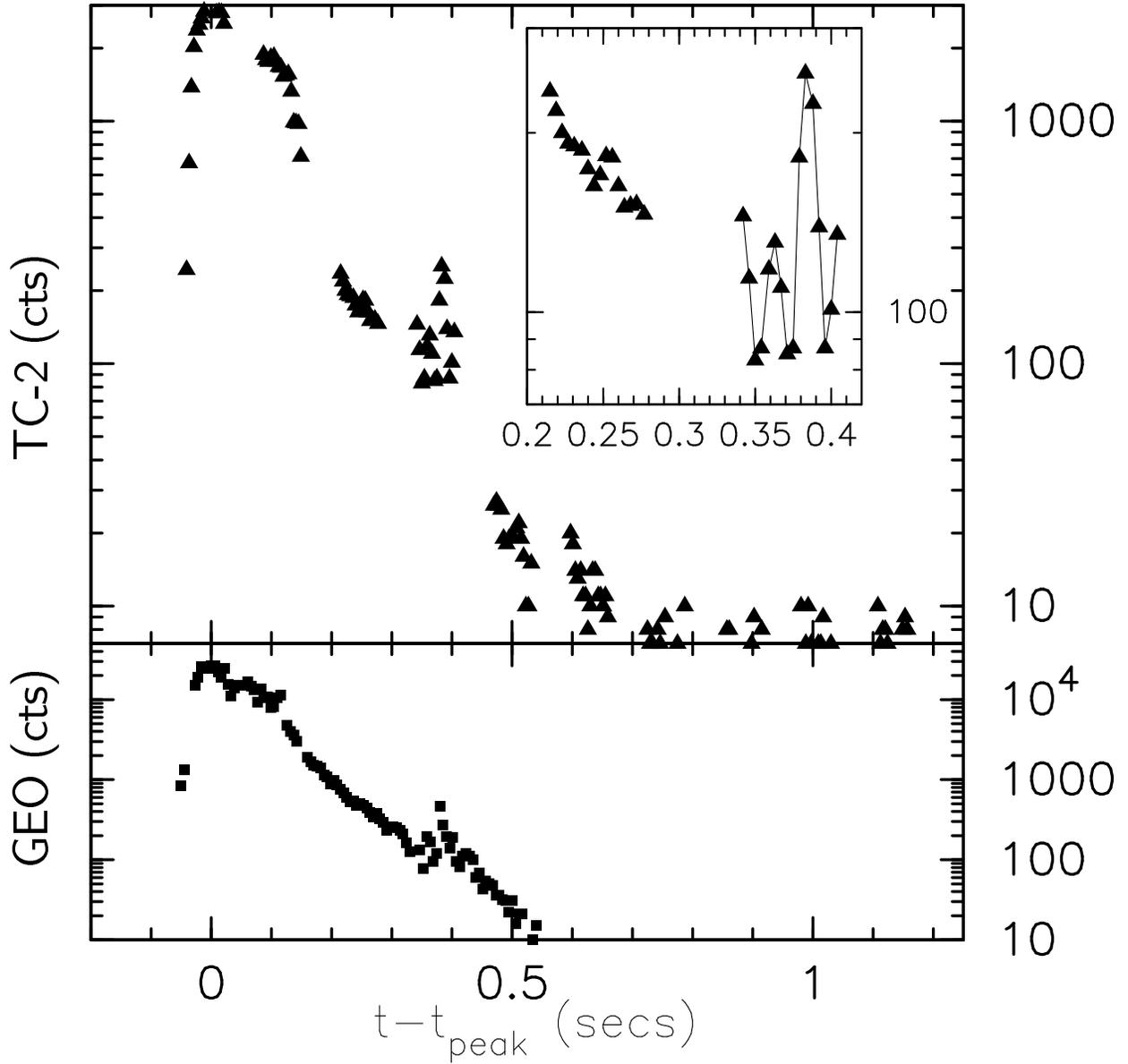}
\caption{(Top) Count rate  returned by the 
PEACE instrument
onboard the Chinese Double Star (TC-2) spacecraft during the intense
$\gamma$-ray event of 27 December 2004. (Bottom)
GEOTAIL count rates  described more fully elsewhere \citep{terasawa++05}.  
A steep rise, peak intensity, shoulder, and
secondary peak are all evident. The signal fades into the background 
count-rate after
$\sim$600ms. The inset shows details of the  secondary
peak at 390ms seen in Figure~\ref{dslightcurve}. This secondary peak compares
well in timing and shape with the feature also seen by BAT/SWIFT
\citep{terasawa++05,palmer++05}.
\label{dslightcurve}}
\end{figure}

\acknowledgements
Some of this work was supported by the UK PPARC. We are grateful to T.\ Terasawa for the
provision of the GEOTAIL data shown in Figure~\ref{dslightcurve} and for many
helpful remarks. We thank G.\ Israel  and C. Thompson for helpful comments and
pre-publication material. Cluster is a project of ESA and NASA. Double Star is a project of the Chinese National Space
Agency together with ESA.

%\bibliographystyle{apj}
%\bibliography{apj-jour,magnetar}

\begin{thebibliography}{20}
\expandafter\ifx\csname natexlab\endcsname\relax\def\natexlab#1{#1}\fi

\bibitem[{Borkowsky {et~al.}(2005)}]{borkowsky++04}
Borkowsky, J., {et~al.} 2005, {GRB} Circular Network, 2920

\bibitem[{Cameron \& Kulkarni(2005)}]{cameron+kulkarni05}
Cameron, P.~B., \& Kulkarni, S. 2005, {GRB} Circular Network, 2928, 2930, 2934,
  2938, 2940, 2946, 2948

\bibitem[{Cline {et~al.}(1980)}]{cline++80}
Cline, T.~L., {et~al.} 1980, \apj, 237, L1

\bibitem[{Duncan \& Thompson(1992)}]{duncan+thompson92}
Duncan, R.~C., \& Thompson, C. 1992, \apj, 392, L9

\bibitem[{Fazakerley {et~al.}(2005)}]{fazakerley++05}
Fazakerley, A.~N., {et~al.} 2005, Annales Geophysicae, submitted

\bibitem[{Feroci {et~al.}(2001)}]{feroci++01}
Feroci, M., {et~al.} 2001, \apj, 549,1021.

\bibitem[{Gaensler {et~al.}(2005)}]{gaensler++05}
Gaensler, B.~M., {et~al.} 2005, {GRB} Circular Network, 2933, 2935, 2943

\bibitem[{Golenetskii {et~al.}(2004)}]{hurley++04}
Golenetskii, S., {et~al.} 2004, {GRB} Circular Network, 2923

\bibitem[{Hurley {et~al.}(2005)}]{hurley++05}
Hurley, K., {et~al.} 2005, \nat, submitted

\bibitem[{Israel {et~al.}(2005)}]{israel++05}
Israel, G., {et~al.} 2005, \apjl, submitted

\bibitem[{Johnstone {et~al.}(1997)}]{johnstone++97}
Johnstone, A.~D., {et~al.} 1997, \ssr, 79, 351

\bibitem[{Lyutikov(2002)}]{lyutikov02}
Lyutikov, M. 2002, \apj, 580, L65

\bibitem[{Lyutikov(2003)}]{lyutikov03}
---. 2003, \mnras, 346, 540

\bibitem[{Mazets {et~al.}(2005)}]{mazets++04}
Mazets, E., {et~al.} 2005, {GRB} Circular Network, 2922

\bibitem[{Mazets {et~al.}(2004)}]{mazets++05}
Mazets, E.~P., {et~al.} 2004, astroph/0502541

\bibitem[{Mereghetti {et~al.}(2005)}]{mereghetti++05}
Mereghetti, S., {et~al.} 2005, \apj, submitted, astroph/0502577

\bibitem[{Palmer {et~al.}(2005)}]{palmer++05}
Palmer, D.~M., {et~al.} 2005, \nat, submitted, astroph/0503030

\bibitem[{Ruderman (1991)}]{ruderman91}
Ruderman, M., 1991, \apj, 382, 587.

\bibitem[{Ruderman {et~al.}(1998)}]{ruderman++98}
Ruderman, M., Zhu, T., \& Chen, K., 1998, \apj, 492, 267.

\bibitem[{Terasawa {et~al.}(2005)}]{terasawa++05}
Terasawa, T., {et~al.} 2005, \nat, submitted, astroph/0502315

\bibitem[{Thompson \& Duncan(1995)}]{thompson+duncan95}
Thompson, C., \& Duncan, R.~C. 1995, \mnras, 275, 255

\bibitem[{Thompson \& Duncan(2001)}]{thompson+duncan01}
---. 2001, \apj, 561, 980

\bibitem[{Thompson {et~al.}(2002)Thompson, Lyutikov, \&
  Kulkarni}]{thompson++02}
Thompson, C., Lyutikov, M., \& Kulkarni, S.~R. 2002, \apj, 574, 332

\end{thebibliography}

\clearpage

\clearpage

\end{document}